\documentstyle[epsf]{kluwer}
\input psfig.sty
\def\gapprox{\;\rlap{\lower 2.5pt            
 \hbox{$\sim$}}\raise 1.5pt\hbox{$>$}\;}       
\def\lapprox{\;\rlap{\lower 2.5pt            
 \hbox{$\sim$}}\raise 1.5pt\hbox{$<$}\;} 
\def\N{\,{\rm I\kern-.20em N}}
\begin{document}

\begin{article}
\begin{opening}
\title{Microflares and hot component in solar active regions}
\author{Arnold O. \surname{Benz}}
\author{Paolo C. \surname{Grigis}}

\institute{Institute of Astronomy, ETH Z\"urich, CH-8092 Zurich, Switzerland}
\runningtitle{Microflares in solar active regions}
\runningauthor{A. O. Benz and P. C. Grigis}

\date{Received:.......  ; accepted :....... }
\begin{abstract} 
Open-shutter RHESSI observations of 3--15 keV X-rays are found to exhibit active region transient brightenings and microflares at a rate of at least 10 per hour occurring even during the periods of lowest solar activity so far in the mission. A thermal component fitted by temperatures of 6--14 MK dominates from 3 keV to about 9 keV, but can be traced up to 14 keV in some cases, and has an average duration of 131($\pm$103) seconds at 7--8 keV. The duration increases with decreasing photon energy. The peak count rate defined by cross-correlation is delayed at low energies. The temperature peaks early in the event and then decreases, whereas the emission measure increases throughout the event. The properties are consistent with thermal conduction dominating the evolution. In some of the bigger events, a second component was found in the 11--14 keV range extending down to 8 keV in some cases. The duration is typically 3 times shorter and ends near the peak time of the thermal component consistent with the Neupert effect of regular flares. Therefore the second component is suggested to be of non-thermal origin, presumably causing the beam-driven evaporation of the first component. The two components can be separated and analyzed in detail for the first time. Low-keV measurements allow a reliable estimate of the energy input by microflares necessary to assess their relevance for coronal heating.
\end{abstract} 
\end{opening}
\section{Introduction}
Plasma of many million degrees and unknown origin has been reported from solar active regions (e.g. Brosius {\it et al.}, 1996). Observers of the full Sun find a power-law distribution for the differential emission measure (DEM) decreasing with temperature. The higher the level of activity, the larger is the coronal emission measure and the flatter its temperature distribution. In the presence of large active regions, it has been observed to extend to temperatures higher that 10 MK (e.g. Peres {\it et al.}, 2000, and references therein), but is not well-observed beyond 10 MK. On top of the power-law distribution, flares produce transient peaks. The DEM of flares has a peak at 6($\pm 1$) MK for A2 flares increasing up to 23($\pm 3$) MK for X1 flares (Feldman {\it et al.}, 1996). At any phase, the flare DEM is quite narrow and covers less than a decade in temperature (Reale, Peres and Orlando, 2001). From observations unresolved in time and space thus emerges the question whether the high temperature plasma apparent outside flares is just the remnant of previous larger flares, is heated by a stationary process or results from a superposition of unresolved small flares.

Small flares in the 3 -- 8 keV range lasting a few minutes have first been observed by HXIS on SMM (Simnet {\it et al.}, 1989). Shimizu (1995) found microflares (active region transient brightenings) in Yohkoh/SXT observations sensitive to 4--7 MK. He reported events having energies as low as a few $10^{26}$ergs and a correlation of peak flux with source size (loop length) and density. Watanabe {\it et al.} (1995) have observed the full Sun in the high-temperature Sulfur XV lines using the BCS instrument on board Yohkoh, indicating brightenings in excess of 10 MK. Note in contrast that the thermal energy content of micro-events in the quiet corona is more than two orders of magnitude smaller, and temperatures of only 1.0--1.6 MK are reported (Krucker {\it et al.}, 1997, and reviewed by Benz and Krucker, 2002).

Lin {\it et al.} (1984 and references therein) have discovered small HXR flares produced by $>$20 keV electrons. Comparing the $>$25 keV channel of LAD/CGRO with microflares observed in the 8--13 keV channel of SPEC/CGRO, Lin, Feffer and Schwartz (2001) reported many low-energy events that have no counterpart at high energies. If interpreted as emission of the same non-thermal electron population, it would suggest that the spectrum steepens or cuts off between 10 and 25 keV. Thus the nature of the low-energy events and their relation to the non-thermal events at high energy remain questionable.

In this paper, we present microflare observations by the RHESSI satellite (Lin {\it et al.}, 2002) at unprecedented spectral and temporal resolution of photons above 3 keV, highly sensitive to the highest temperatures reported for active regions as well as to the lowest energies of non-thermal photons expected from microflares. First RHESSI results on microflare imaging are presented by Krucker {\it et al.} (2002). Here we explore the spectral and temporal characteristics of microflares and concentrate on the question whether the emissions are thermal or non-thermal.
\section{Observations and Results}

The Reuven Ramaty High Energy Solar Spectroscopic Imager (RHESSI) was launched into orbit on 5 February 2002. The germanium detectors register photons in the energy range from 3 keV to 17 MeV with 1 keV resolution at low energies (Smith {\it et al.}, 2002). Imaging is achieved by nine absorbing grids modulating by satellite rotation. It allows to reconstruct the full Sun image with a resolution of 2$''$ at low energies (Hurford {\it et al.}, 2002). At low count rates, the spatial resolution is degraded (Saint-Hilaire and Benz 2002). RHESSI observes the Sun full time interrupted only by satellite night and the South Atlantic Anomaly. Energetic electrons sometimes cause spurious counts when the satellite is at high geographical latitudes. We use standard RHESSI software and calibrations updated to mid-August 2002.

As the energy, time and detector information of each photon is registered, the large number of low-energy photons during large flares causes pile-up in the detector. To bring it down to a manageable level, one or two shutters are usually put in front of the detectors to reduce the count rate at low energies.  Early in the mission, the shutters were open only in relatively quiet times, when no large flares were expected. 

Here we have selected RHESSI observing times when no shutter was absorbing the low-energy photons. During these times the full flux of low-energy photons is available. Thus these times yield maximum sensitivity for low-energy studies. We have found some 50 open-shutter orbits in the months of March, April and May 2002. Uninterrupted intervals at low solar activity and low background were selected for further analysis. In particular, we required {\sl (i)} a count rate less than 600 cts per second and per detector in the 3--12 keV range and {\sl (ii)} a constant background in the 100--300 keV range. A total of seven intervals satisfied these requirements. The GOES level in all of them is below C, and all flares are below GOES class A9 after subtracting the background.

\subsection{Light Curves}

Figure 1 displays RHESSI full Sun observations at low energies during one-orbit intervals. The low peak count rates justify neglecting the pile-up effect of photons in the detectors. The photons have been integrated in one keV bins that we will refer to as channels. The 10--11 and 11--12 keV channels have a lower signal-to-noise ratio because of an instrumental background line feature. Also displayed in Fig. 1 are the GOES 8 observations in the low-energy band, extending the energy coverage below the RHESSI range. 
\begin{figure*}
\begin{center}
\leavevmode
\mbox{\hspace{0.0cm}\epsfxsize=12cm
\epsffile{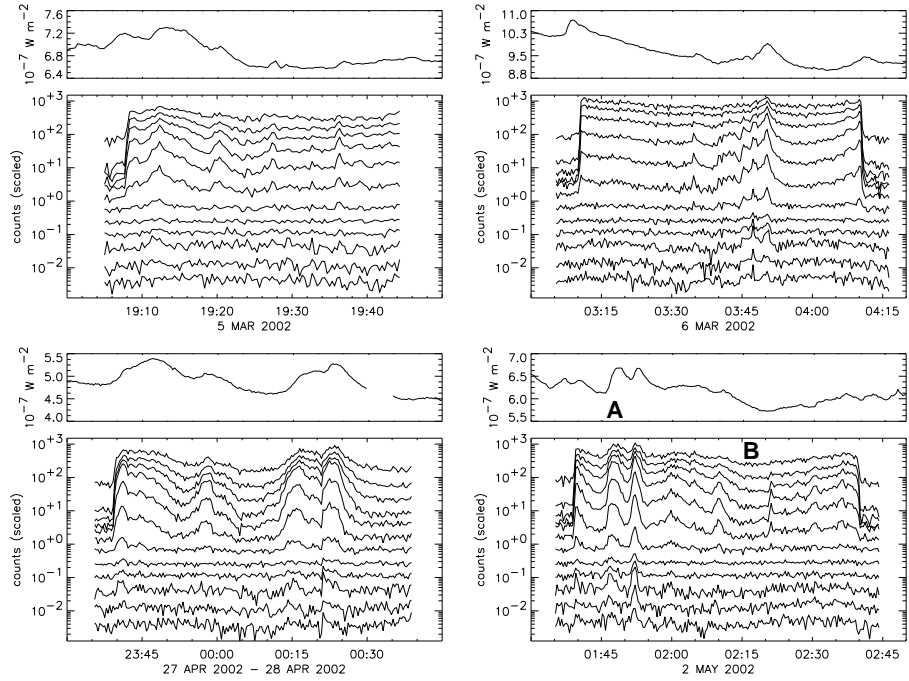}}
\end{center}
\caption[]{{\sl Top:} Light curve observed by GOES 8 in the 1--8 \AA\ band (1.6--12.4 keV) of soft X-rays. {\sl Bottom:} RHESSI light curves in one keV channels from 3--4 keV (top) to 14--15 keV (bottom). Each channel is multiplied such that it does not overlap with others. The time resolution is 20 s. {\sl a)} Sun light starts at 19:08 UT. {\sl b)} Sun light starts at 03:11 UT and ends at 04:10 UT. {\sl c)} Sun light starts at 23:40 UT. {\sl d)} Sun light starts at 01:40 UT and ends at 02:40 UT. }  
  \label{fig:1}
\end{figure*}

Visual inspection of Fig. 1 yields many events per orbit. There is no time of stationary flux. During the total selected observing time of 373 minutes, we found 64 events, thus 10.3 per hour on average in one solar hemisphere. Lin {\it et al.}(1984) reported 10 h$^{-1}$, Lin {\it et al.}(2001) 5.5 h$^{-1}$, and Shimizu (1995) 26 h$^{-1}$, all of them observing during higher levels of solar activity. 

Some general properties follow immediately from Fig.1:
\vskip-0.5cm
\begin {itemize}
\item The largest variability occurs in the channels between 5--9 keV.
\vskip-0.5cm
\item Some events in the 6--9 keV channels are not visible in the GOES light curve, being dominated by photons $<$ 3 keV (see e.g. event after B in Fig. 1d. GOES shows no event larger than A0.1 above background).
\vskip-0.5cm
\item The emission above about 9 keV peaks generally before the emission at lower energies.
\vskip-0.5cm
\item The event duration decreases continuously from 1.6 keV (GOES) to 15 keV.
\vskip-0.5cm
\item In most events there is no emission perceivable beyond about 12~keV.
\end {itemize}

\subsection{Photon Energy Spectra}

\begin{figure}
\begin{center}
\leavevmode
 \mbox{\hspace{0.0cm}\epsfxsize=11.5cm
\epsffile{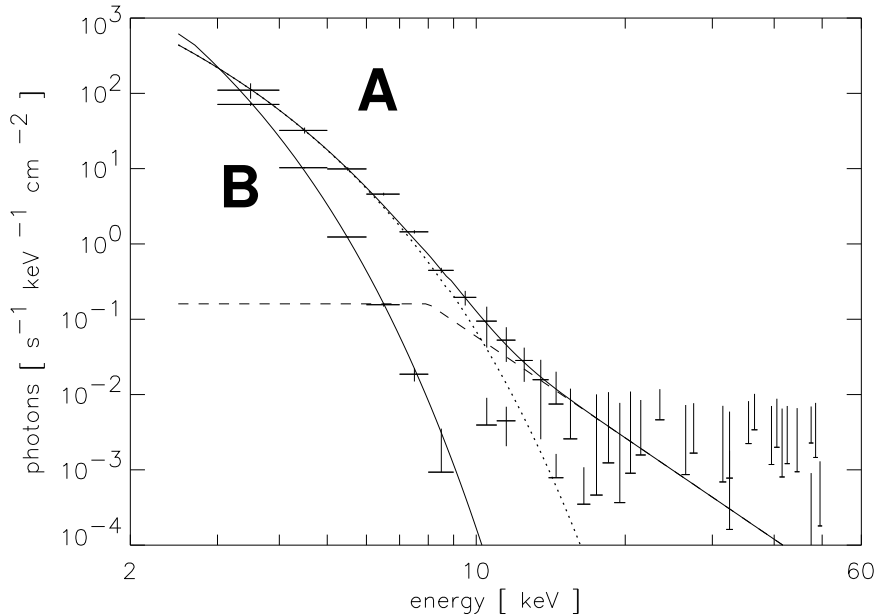}}
\end{center}
\caption[]{Spectrum of event A and background B on 2-May-2002 (Fig.1d) observed with RHESSI. The event is fitted with a thermal contribution (dotted) and a non-thermal component (dashed) having a break point at 8 keV. The background is fitted with only a thermal contribution.}  
  \label{fig:2}
\end{figure}

Two spectra are shown in Fig.2. The data were calibrated using the full detector response matrix. The first spectrum was integrated during event A shown in Fig.1d from 01:46:40 to 01:47:40 UT, the maximum phase at the higher energies. The total background, taken at the time of minimum flux before the event, was subtracted. Only the front detectors 1, 3, 5, and 9 have been used, having the lowest dropout rates in the time interval. Data dropouts are suspected to be caused by heavy cosmic rays interfering with the RHESSI electronics. Although the software corrects for the deadtime due to these data gaps, they introduce additional noise. 

At low energies the spectrum fits well with an isothermal plasma at a temperature of 12.1($\pm$ 0.4) MK and an emission measure of 7($\pm 2)\times 10^{45}$cm$^{-3}$.  The best fit is shown in Fig. 2, where all normalized residuals are below 0.4. The errors originate mainly from the choices of the background and the low-energy break point for non-thermal emission. As the power-law distribution of the non-thermal component must turn over at low energies, it was approximated by a constant below some energy. The quality of the fit is not sensitive the break energy; best fits are between 6 and 9 keV. In the case of Fig. 2 different background choices changed temperatures in the range of $\pm 0.8$ MK, and different choices for the low-energy break for non-thermal emission produced a range of $\pm 0.7$ MK. The standard error was estimated by taking a third of the rms sum of the two. Note that the error in temperature and emission measure are anti-correlated.

At the time of writing, the energies below 5 keV are not yet fully calibrated. This can be noted in the fact that single collimators give a range of temperature values of $\pm$1.7 MK if using only the lowest 3 keV. The scatter decreases by a factor of 3 for taking a larger range. At 6--7 keV, and possibly also at 7--8 keV, the flux is often enhanced. This cannot be reduced substantially by the choice of the break point. The enhancement could be caused by the 6.7 keV iron line emissions not included in the fit. Therefore we did not use 6--8 keV for fitting. Instead the break point was put at 8 keV in accordance with other information (Figs. 4 and 5).

In the energy range of 10 -- 15 keV, the isothermal model does not fit the observations of microflare A (Fig.2). A power-law photon distribution was therefore added. Its fitted slope depends on the break point and has an exponent in the range -4.5 $\lapprox \gamma \lapprox$ -5.7. For the choice of the break point at 8 keV, the best fit yields -4.5. The spectrum disappears in the background beyond 15 keV.

Instead of a power-law, the 10 -- 15 keV spectrum could also be fitted by a second thermal component with higher temperature and smaller emission measure. The best fit is achieved for a temperature of 24.7 MK and an emission measure of $0.14\times 10^{45}$cm$^{-3}$. The contribution of the hot kernel would not be discernible below 8 keV and fit all other constraints. 

The thermal energy content of an isothermal plasma is

\begin{equation}
E_{th} \approx 3\sqrt{{\cal{M}}V} k_B T
\end{equation}
It is assumed in Eq.(1) that the electron and ion densities are about equal and constant in the volume $V$. $\cal{M}$ is the emission measure, and $k_B$ is the Boltzmann constant. Putting in the observed values, we find for the thermal content of the microflare (index A) and of the hot kernel (index k)

\begin{equation}
E_{th}^A \approx 4.2\times 10^{14}\sqrt{V_A} \ \ \ \ \ {\rm [erg]}\ \ ,
\end{equation}
\begin{equation}
E_{th}^{k} \approx 1.2\times 10^{14}\sqrt{V_{k}} \ \ \ \ \ {\rm [erg]}\ \ ,
\end{equation}
Thus the hot kernel would contain less energy than the main component for $V_k < V_A$.

The spectrum at the time of the lowest count rate in the orbit (marked B in Fig.1d) is also displayed in Figure 2. It is the time interval from 02:14:40 -- 02:19:56 UT, when no microflare was observed and the most recent one had occurred more than 4 minutes before (Fig.1d). For this case, the background interpolated from the satellite's night time before and after the observations was subtracted. An isothermal model having a temperature of 6.4 MK fits well. No non-thermal or hot kernel component is above the background level. The emission measure of the hemisphere is measured $45\times 10^{45}$cm$^{-3}$, and its thermal energy content is

\begin{equation}
E_{th}^{B} \approx 5.6\times 10^{14}\sqrt{V_B} \ \ \ \ \ {\rm [erg]}\ \ .
\end{equation}
The fact that Eqs.(2) and (5) have similar factors does not mean that $E_{\rm th}^A \approx E_{\rm th}^B$ since probably $V_B \gg V_A$.

\begin{figure}
\begin{center}
\leavevmode
 \mbox{\hspace{0.0cm}\epsfxsize=11.5cm
\epsffile{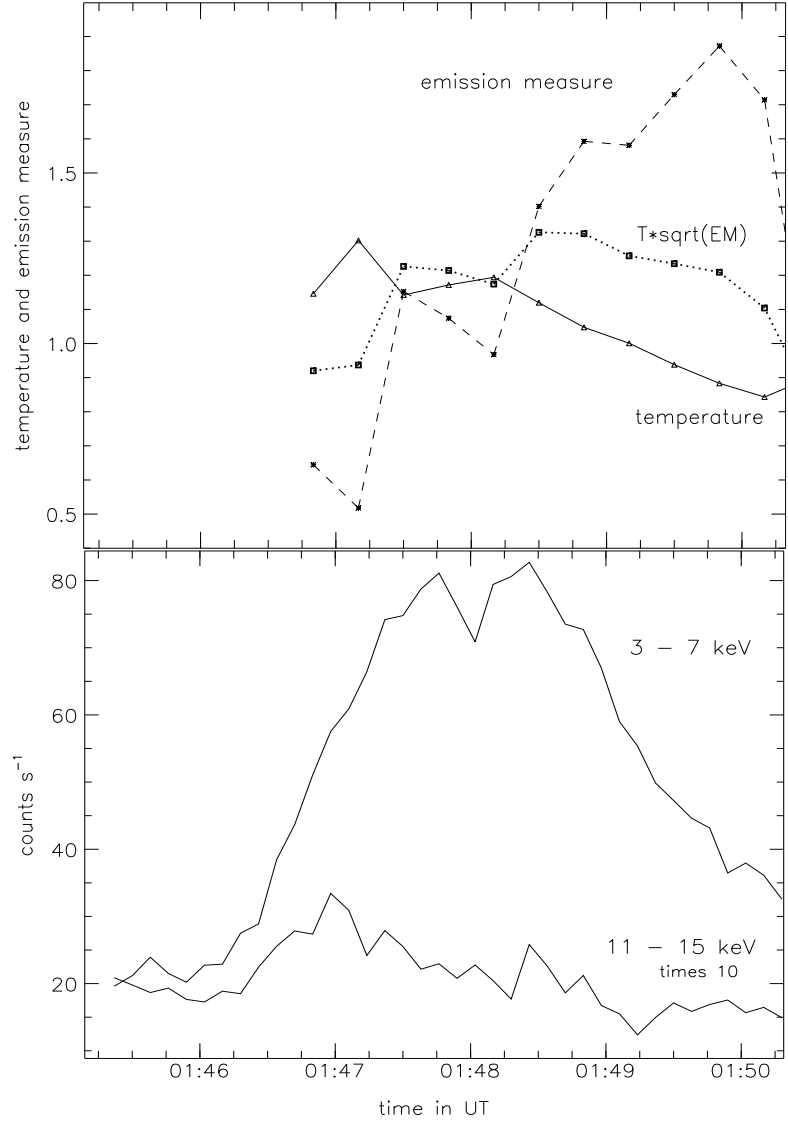}}
\end{center}
\caption[]{{\sl Top:} Temperature (solid) in units of 10 MK, emission measure (dashed) in $10^{-46}$cm$^{-3}$ and the product $T\sqrt{{\cal{M}}}$ (dotted) of event A in Fig.1d (thermal component). {\sl Bottom:} Light curve of event A in the energy bands 3--8 keV and 11--15 keV (multiplied by factor of 10). The integration time is 8 s, the data are calibrated by the diagonal matrix elements.}
  \label{fig:3}
\end{figure}

Microflare A can be imaged with RHESSI collimators 3 to 9. The microflare took place in NOAA Region 9932 (see image and movie on CD) and had a nominal FWHM size of 8$''$$\times$11$''$. It is close to the resolution limit, thus $V_A \lapprox 3\times 10^{26}$cm$^3$. Eq.(2) yields an upper limit on the thermal energy content of $7.2\times 10^{27}$erg. A lower limit for the electron density of $4.8\times 10^{9}$cm$^{-3}$ is derived from emission measure and volume assuming a filling factor of unity. Microflare imaging is studied in detail by Krucker {\it et al.}(2002).

The relation between the temperature of the thermal component and the emission measure is further explored in Fig. 3. Temperature and emission measure have been determined identically to Fig. 2. Using the same background and break point that produce similar systematic errors, the remaining relative error between measurements is estimated about $\pm $0.2 MK, as determined from different fits. The temperature maximum occurs between peak fluxes in 12--15 keV and 3--7 keV. The product $T\sqrt{{\cal M}}$ is proportional to the energy content if the volume remains constant. The product increases initially until about 01:48:30 s, when the 12--15 keV count rate has dropped to the half-maximum level, and then remains approximately constant throughout the microflare.

\subsection{Low and High Energy Photons}

The light curves integrated from 3--7 keV (low energy) and 12--15 keV (high energy) are shown in Fig. 3. The low-energy band peaks at 01:48:10($\pm$20) UT. The high-energy count rate reaches the preflare background level of 1.8 counts s$^{-1}$ at 01:48:40($\pm$20) UT. Thus the two light curves exhibit the Neupert effect (Neupert, 1968) known in regular flares. It is generally interpreted there as the signatures of different emissions relating to a thermal and a non-thermal electron population in the scenario of electron-beam driven evaporation.

The high-energy band peaks 80 seconds before the low-energy band. The FWHP duration at 12--15 keV is only 45 s, whereas it is 148 s at 3--5 keV.

\begin{figure}
\begin{center}
\leavevmode
\mbox{\hspace{0.0cm}\epsfxsize=11.5cm
\epsffile{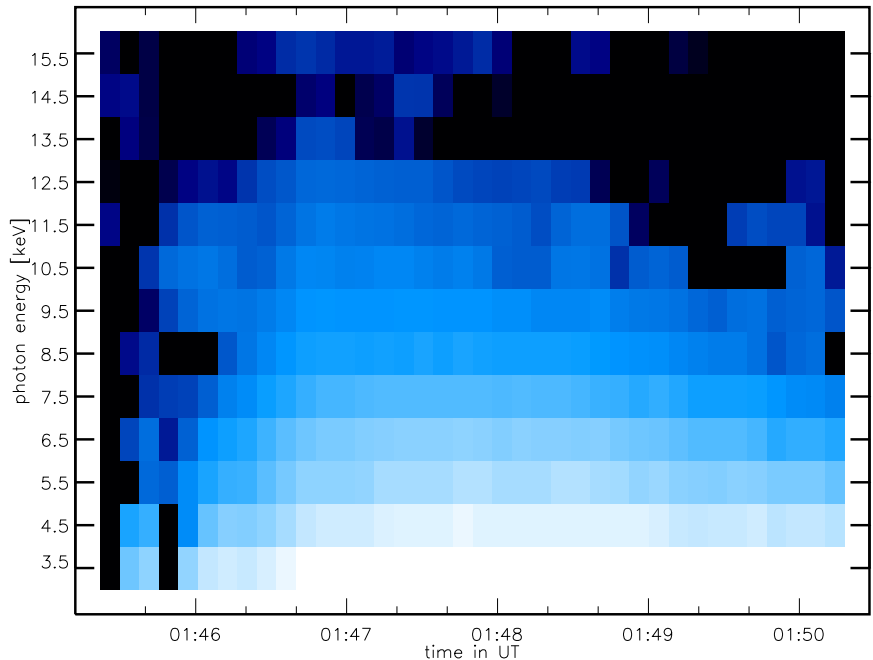}}
\end{center}
\caption[]{Spectrogram observed by RHESSI of event A in Fig.1d. The energy channels (1 keV) and time bins (8 s) are presented by pixels. A logarithmic scale is applied, bright meaning enhanced photon flux.}  
  \label{fig:4}
\end{figure}

Further differences between low and high-energy photons appear in the spectrogram (Fig.4). The data are calibrated using a diagonal response matrix. The high-energy peak at 01:46:50 UT is clearly visible. It takes a straight vertical shape emerging at about 8 keV out of the thermal part. The latter has a triangular shape with an apex at about  01:47:25 UT and 12 keV.
%
\begin{figure}
\begin{center}
\leavevmode
\mbox{\hspace{0.0cm}\epsfxsize=11.5cm
\epsffile{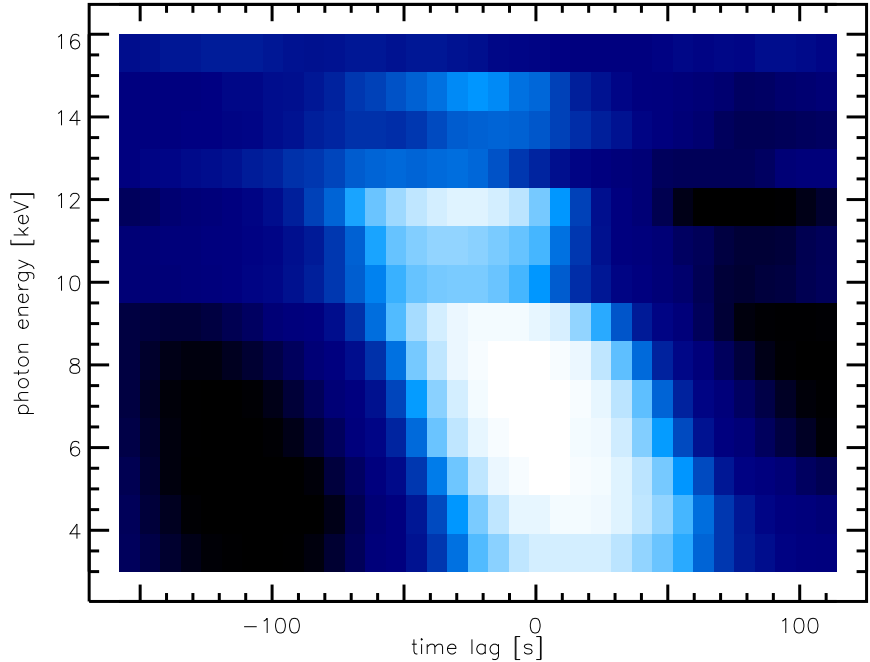}}
\end{center}
\caption[]{Correlogram of Fig.4 showing the cross-correlation of relative to the 7--8 keV channel. Bright pixels represent high correlation coefficients.}  
  \label{fig:5}
\end{figure}

The timing of photon fluxes at different energies was studied also by cross-correlation. The light curves at 4 s resolution were each cross-correlated relative to the most variable channel (7--8 keV). The derived cross-correlation coefficients revealed  a local maximum at zero lag even at energies $>$20 keV, where no solar photons were detected above background. As the zero-lag cross-correlation peak also exists at satellite night, it must be an instrumental effect. Most likely it is due to data dropouts, except for the auto-correlation in the 7--8 keV channel. The effect was eliminated by interpolation between the neighboring lags and integration to 8s. 

Figure 5 depicts the result of the cross-correlation in spectrogram form (correlogram). Channels with large background, such as 10--11 and 11--12 keV, have higher noise and yield lower cross-correlation coefficients. Again, the difference between low and high energy components appears. Most relevant is the change in drift rate of the cross-correlation peak. At energies higher than 9 keV, the peak does not drift but forms a vertical structure at a constant lag of -25s. Below 9 keV, the peak drifts to larger lags (from upper left to lower right), reaching about +25s at 3--4 keV.

\begin{figure}
\begin{center}
\leavevmode
\mbox{\hspace{0.0cm}\epsfxsize=11.5cm
\epsffile{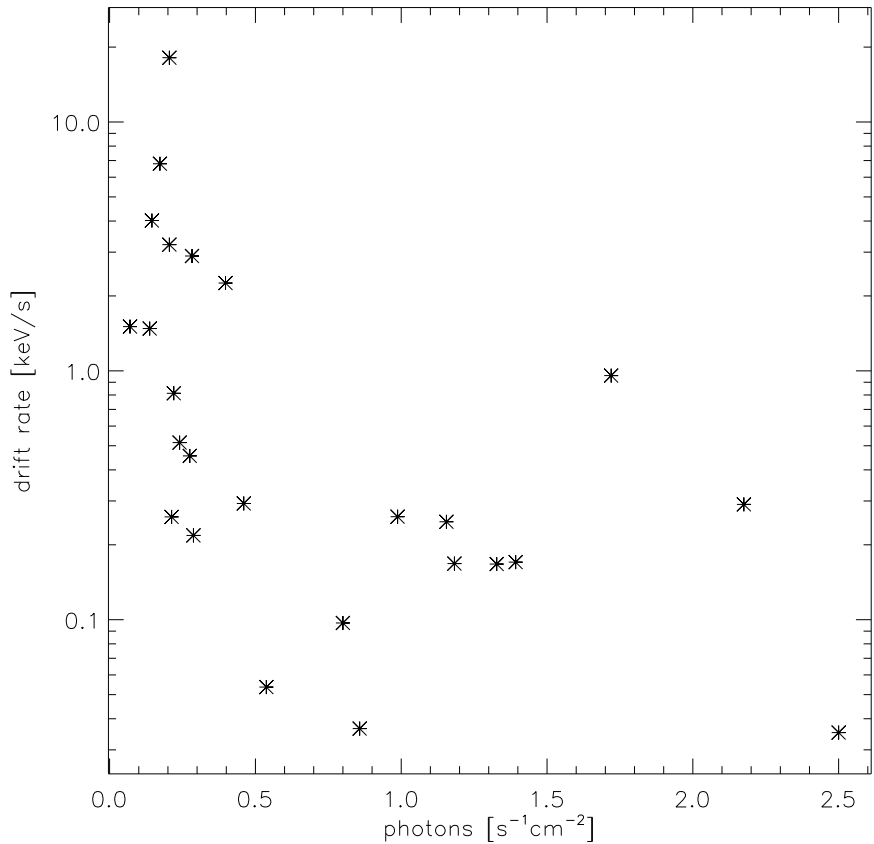}}
\end{center}
\caption[]{Drift rate of the light curves of the 1 keV channels between 3 and 9 keV vs. peak photon flux in the 7--8 keV channel.}  
  \label{fig:6}
\end{figure}

\subsection{Low-energy Component}

The 25 largest microflares in the sample have been analyzed similarly to the above analysis. Their average peak flux in the 7--8 keV band above the preceding minimum is 0.7 photons s$^{-1}$ cm$^{-2}$. The average FWHM duration is 131($\pm$103) seconds (standard deviation), and the median value is 114 s. We found no correlation between peak flux and duration in agreement with Shimizu's (1995) measurements at lower energies.

A relation between peak flux and absolute drift rate of the cross-correlation peak is apparent from Fig. 6. The drift rate was determined from the cross-correlation coefficients of the 1-keV channels in the 3--9 keV range by linear (Fig.5) of the peak values in energy and time. The peaks in each channel are determined from a boxcar smoothing of 40 s and spline interpolation. Spline and regression allow a higher resolution and the determination of larger drift rates than from the pixels in the spectrogram. Nevertheless, the error increases enormously at drift rates beyond about 2 keV/s. Large microflares have slower absolute drift rate, which is usually negative (peak drifting from high to low energies). Some smaller microflares have large absolute drift. In one case out of 25 it was found positive, but the error range includes negative values. The anti-correlation between photon flux and drift rate is more prominent for events during the same orbit. 

The spectrum of X-ray microflares at energies in the 3--9 keV range is well modeled by an isothermal plasma as shown in the example of the previous section. The best fitting temperatures in the 25 largest and well isolated events are in the range 10.2 -- 14.7 MK, with an average value of 12.0($\pm$1.2)  MK. This is considerably more than reported previously (Shimizu, 1995; Feldman {\it et al.}, 1996). The difference seems to originate from Yohkoh/SXT and GOES favoring  the 4--7 MK range, but RHESSI observing the most energetic thermal photons and thus emphasizing the highest temperatures. This is supported by GOES observations of the microflare of Fig. 2 (GOES class A5) where the two GOES channels having lower energy bounds of 1.6 and 3.2 keV indicate a temperature of only 7$\pm$1 MK (background subtracted).

\subsection{High-energy Component}

The second component is best observable in 11--14 keV photons. As a test of its presence, the $<$8 keV spectrum was fitted by a thermal model, and the 12--13 keV channel compared to the fit. In 6 of the 25 largest microflares this channel was clearly above the thermal, 2 more cases were doubtful. In 4 cases, all at relatively high temperature, the 12--13 keV flux was within the error bars consistent with the thermal model. In the rest, 13 cases, the 12--13 keV flux was not significantly above background and could not be classified. 

\section{Discussion}

The existence of a second, non-thermal component at high energies in microflares is not obvious. We have searched for it in the following ways:

1. The {\sl spectrum} (Fig.2) of microflares below 10 keV is dominated by a thermal component with a temperature of the order of 10 MK. In 24\% of the cases an enhanced high-energy tail is observed extending up to 15 keV. It could also be interpreted by a hot thermal kernel.

2. The {\sl spectrogram} (Fig.4) shows more clearly the difference between the two components: {\sl (i)} The low-energy thermal component having a triangular shape with a long decay at the lowest photon energies and {\sl (ii)} a short vertical structure extending to higher energies in the rise phase of the former component. 

3. The drift of the {\sl cross-correlation peak} at low energies is a combination of two effects: {\sl (i)} The maximum of the light curve is more and more delayed the lower the energy, and {\sl (ii)} the shape of the microflare becomes asymmetric at lower energies having a longer decay time than rise time (Fig.3). Cross-correlation is thus also a method to identify the two components and to separate them in energy. 

There are two major indications for a "non-thermal" origin of the high-energy component:
\begin{itemize}
\item A hot kernel of thermally emitting plasma also fits the 9--14 keV spectrum, but would produce a drift in the correlogram during cooling that is not observed in the second component. 
\item The second component has shorter duration, occurs in the early phase of the thermal component and vanishes at the peak flux of the thermal component. It resembles the non-thermal component of regular flares in its relation to the first component known as the Neupert effect. The second component thus exhibits the characteristics of thick target bremsstrahlung of an electron beam. 
\end{itemize}
Therefore, we interpret the second component of microflares as non-thermal emission. 

An important result is the increase of the emission measure throughout the event (Fig.3). It indicates that the amount of high-temperature material augments (excluding the unrealistic possibility of compression). At the same time the temperature decreases. A surprising characteristic is the flat top of the product 
$T\sqrt{{\cal{M}}}$ from the time of the temperature peak early in the flare until the late decay phase. Under the assumption of constant volume, the flat top suggests a constant thermal energy content 

Thermal conduction could explain such a behavior if radiative energy loss can be ignored. The thermal component may start with a high temperature and a small mass. Thermal conduction then adds more material to the temperature range observed, but decreases the mean temperature at the same time. It is consistent with the observation of the thermal component, having a later peak and longer duration at low energies.

Conduction can also explain the drift of the cross-correlation (Fig.5). As the temperature decreases, the count rate in the higher energy channels diminishes more rapidly than at low energies where the increasing emission measure even delays the maximum flux. The smaller the flare size, the faster this takes place. Thus the cross-correlation peak drifts the faster, the smaller the flare. Considering the previously reported relation between flare size and peak flux (see Introduction), this may explain the observed inverse relation between drift rate and peak flux (Fig.6).

Conduction models have been proposed to interpret a Neupert-like relationship between higher and lower energy photons (reviewed in Dennis and Schwartz, 1989). Although the second component of microflares at high energies could be fitted by the thermal emission of a hot kernel, it would not contain enough energy to allow for conductive heating of the first component. Therefore, conduction may explain the evolution of the first component, but cannot explain the relation between the first and second component. 

We propose here that the flare duration at 3--9 keV (thermal component) is not controlled by radiation loss, but by thermal conduction distributing the energy to a larger mass reservoir at lower temperature.

\section{Conclusions}
The X-ray spectrum from 3 -- 15 keV of solar microflares in active regions has been observed with high energy resolution by the RHESSI satellite. The thermal component can be identified in all microflare and traced up to some 12 keV in the spectrogram. It can be well separated from a second component at higher energies, not visible in all events, that peaks simultaneously at all energies, has short duration and occurs during the rise phase of the first component. The second component can be traced in energy down to about 8 keV. The existence of two components excludes a previous interpretation (see Introduction) as one non-thermal electron population having a cut-off distribution in energy.

The similarities of microflares to regular flares suggest identical physical processes and similar plasma conditions. The properties of the thermal and non-thermal components are consistent with the standard flare scenario of precipitating electrons heating cold material to flare temperature. The collisional energy loss of beaming electrons causes bremsstrahlung emission observable in the 9--15 keV range and heats cold material to emit thermally at $<$12 keV. 

High spectral resolution at the transition between thermal and non-thermal emission has been found to be useful for studying the role of energy deposition by beams and thermal conduction. No-shutter observations by RHESSI are well suited to further quantify these results. This will lead to firm estimates on the various forms of energies involved in microflares and determine their role in heating the active region corona.

\begin{acknowledgements}
RHESSI is a NASA Small Explorer Mission with an important hardware contribution by the Paul-Scherrer Institute (Villigen, Switzerland). The HESSI Experimental Data Center is supported by ETH Zurich. Many people have contributed to the successful launch and operation of RHESSI. We thank all of them, and in particular Pascal Saint-Hilaire, for help and explanations, Hugh Hudson and S\"am Krucker for helpful comments. Oliver Trachsel has developed the software for the drift rate measurement. The RHESSI work at ETH Zurich is partially supported by the Swiss National Science Foundation (grant nr. 20-67995.02).
\end{acknowledgements}

\end{article}
\end{document}